\documentclass[preprint]{aastex}
\usepackage{multirow,natbib,epsf,epsfig}

\newcommand{\kms}{km\ s$^{-1}$}

\newcommand{\Msun}{$M_{\sun}$}

\slugcomment{Accepted for Publication in ApJ}

\shorttitle{Hot Disks \& Delayed Star Formation}
\shortauthors{Sheth et al.}

\begin{document}


\title{Hot Disks \& Delayed Bar Formation}

\author{Kartik Sheth \altaffilmark{1}, Jason Melbourne
  \altaffilmark{2}, Debra Meloy
  Elmegreen \altaffilmark{3}, Bruce G. Elmegreen\altaffilmark{4}, E.
  Athanassoula \altaffilmark{5}, Roberto G. Abraham \altaffilmark{6},
  Ben Weiner \altaffilmark{7}}

\altaffiltext{1}{National Radio Astronomy Observatory, 520 Edgemont Road, Charlottesville, VA 22903}
\altaffiltext{2}{California Institute of Technology, MC 105-24, 1200 East California Boulevard, Pasadena, CA 91125}
\altaffiltext{3}{Department of Physics and Astronomy, Vassar College, 124 Raymond Avenue,  Poughkeepsie, NY 12604}
\altaffiltext{4}{IBM T. J. Watson Center, P.O. Box 218, Yorktown Heights, NY 10598}
\altaffiltext{5}{Aix Marseille Universite, CNRS, LAM (Laboratoire d'Astrophysique de Marseille) UMR 7326, 13388, Marseille, France}
\altaffiltext{6}{Canadian Institute for Theoretical Astrophysics, Mclennan Labs, University of Toronto, 60 St. George St, Room 1403, Toronto, ON M5S 3H8, Canada}
\altaffiltext{7}{Department of Astronomy/Steward Observatory, 933 North Cherry Avenue, Tucson, AZ 85721-0065}

\begin{abstract}

We present observational evidence for the inhibition of bar formation in dispersion-dominated (dynamically hot) galaxies by studying the relationship between galactic structure and host galaxy kinematics in a sample of 257 galaxies between 0.1 $<$ z $\leq$ 0.84 from the All-Wavelength Extended Groth Strip International Survey (AEGIS) and the Deep Extragalactic Evolutionary Probe 2 (DEEP2) survey.  We find that bars are preferentially found in galaxies that are massive and dynamically cold (rotation-dominated) and on the stellar Tully-Fisher relationship, as is the case for barred spirals in the local Universe.  The data provide at least one explanation for the steep ($\times$3) decline in the overall bar fraction from z=0 to z=0.84 in L$^*$ and brighter disks seen in previous studies.  The decline in the bar fraction at high redshift is almost exclusively in the lower mass (10 $<$ log M$_{*}$(\Msun)$<$ 11), later-type and bluer galaxies. A proposed explanation for this  ``downsizing'' of the bar formation / stellar structure formation is that the lower mass galaxies may not form bars because they could be dynamically hotter than more massive systems from the increased turbulence of accreting gas,  elevated star formation, and/or increased interaction/merger rate at higher redshifts.  The evidence presented here provides observational support for this hypothesis.  However, the data also show that not every disk galaxy that is massive and cold has a stellar bar, suggesting that mass and dynamic coldness of a disk are necessary but not sufficient conditions for bar formation -- a secondary process, perhaps the interaction history between the dark matter halo and the baryonic matter,  may play an important role in bar formation.  

\end{abstract}


\section{Background}\label{intro}

The presence of galactic structures such as bars is as an  important signpost in the evolution of a galaxy disk.  Analystical work and simulations have shown that, once a galaxy disk is sufficiently massive and dynamically cold, the formation of a stellar bar is relatively fast ($\sim$hundred million years) (e.g., \citealt{hohl71, kalnajs72, ostriker73, sellwood93, athanassoula02, athanassoula03, heller07}).  But bar formation can be delayed either by an {\em initially} dominant dark matter (DM) halo and/or a dynamically hot/dispersion-dominated disk \footnote{We use the terms dynamically hot and dispersion-dominated interchangeably throughout the paper.} \citep{athanassoula86}\   \\

An {\em initially} dominant DM halo strongly impacts the time scale for bar formation delaying the onset of the bar instability \citep{athanassoula02}.  The bar that ultimately forms in such a system is stronger than a bar that would form in an otherwise non-DM dominated galaxy because the DM  halo acts as an efficient sink of angular momentum and energy for baryons, which are redistributed to form the bar.  As the bar grows it pushes material inwards so that the baryonic matter can become the dominant mass component in the inner parts of galaxies \citep{athanassoula02b}.  
Simulations also show that a dynamically hot disk delays bar formation \citep{athanassoula86, athanassoula03} because when random motions of stars in a disk have a higher amplitude than rotational ordered motions, the bar instability cannot grow quickly.  This may even push the bar formation time scale beyond a Hubble time.  \\ 



A recent COSMOS study of over two thousand L$^*$ and brighter, face-on ($i < $ 65$^{\circ}$) disk galaxies showed that the overall bar fraction (f$_{bar} $ = total number of barred galaxies divided by the total number of disk galaxies) in disk galaxies declines sharply from f$_{bar} \sim$ 0.65 at z=0  to f$_{bar}$ $<$ 0.2 at z=0.84 \citep{sheth08}.  It is crucial to note that the COSMOS sample is a complete sample only for disks with stellar masses, M$_* > $ 10$^{10}$\Msun; the published results of the bar fraction evolution apply {\em only} to this mass range \citep{sheth08, cameron10}.  Therefore, studies with samples of lower mass galaxies (M$_* < $ 10$^{10}$\Msun, such as those typically done for nearby galaxies (e.g.,  using SDSS)) are not directly comparable to high redshift studies.  

The evolution of the bar fraction with redshift is not uniform across all disk galaxies.  As a function of redshift, f$_{bar}$ is strongly correlated  with the host galaxy mass, color and bulge dominance (see Figures 2--5 in \citet{sheth08}).  The most massive stellar disks (M$_* \ge $10$^{11}$\Msun), which are also redder and have a larger bulge, already had f$_{bar} >$ 0.5 at z$\sim$0.8, nearly their present-day value of their bar fraction.  In sharp contrast, the lower stellar mass systems (M$\sim$10$^{10}$ \Msun),  had f$_{bar} \ll $ 0.2 at z$\sim$0.8.  Over the last 7 Gyr, the lower mass galaxies have evolved the fastest, increasing their bar fraction by more than a factor of three, to their present day value of f$_{bar} \sim$ 0.65.  This behavior is another form of ''downsizing" \citep{cowie96}.

The dynamics of high redshift disks has been a hot topic of study in recent years (e.g., \citealt{kassin07, schreiber09,cresci09,lorenzo09,davies11,miller11, kassin12}).  At high redshifts (z$\ge$2) there is evidence for both rotation- and  dispersion-dominated disks (e.g., \citep{law09, cresci09,wright09}), although the evolution of the disk kinematics and assembly is not well-understood. The dynamics of a galaxy  must change as it acquires mass, undergoes interactions/mergers and forms stars.  In this paper we seek to understand how the disk dynamics are influencing the formation of bars.   \\

In a 2007 study of over $\sim$500 galaxies from 0.2$<$z$<$1.2, \citet{kassin07} found that major-mergers,  disturbed and compact systems are preferentially off the stellar mass Tully-Fisher (TF) relationship towards lower rotational velocities.  In contrast for the local Universe, \citet{barton01}, who examined 90 close pairs, found that only eight scattered off the TF relationship.  \citep{kannappan02} analyzed the residuals in the TF for a wide variety of galaxy morphologies and environments from the Nearby Field Galaxy Survey and found that the scatter in the TF did not change once corrections for dust extinction and star formation were applied, although the scatter did increase for non-spiral galaxies.  They also found that dwarf galaxies did not follow the TF with dwarfs scattering on both sides of the TF (see \citealt{kannappan02} for an in-depth discussion).  At high redshifts,  Figure 1 of \citet{kassin07} suggests that more early type spirals (blue squares in their Figure 1) are on the classical stellar TF relationship compared to late-type/irregular spirals.  It appears that over time, more and more of late-type/irregular galaxies arrive onto  the stellar-TF.   A different study of disk-like galaxies by \citet{miller11} has argued that there is no significant evolution in the stellar mass TF relationship to z$\sim$1, although there is an evolutionary trend in the B-band TF.  While the precise evolution of the stellar TF is not known, we make use of the existing measurements of disk properties (mass, rotational velocity and velocity dispersion) from \citet{kassin07} and compare these for different types of galaxies.

\section{Defining the Galaxy Sample For Classification of Galactic Structure} \label{sample}
\begin{figure}[ht!]
\epsscale{0.8}
\plotone{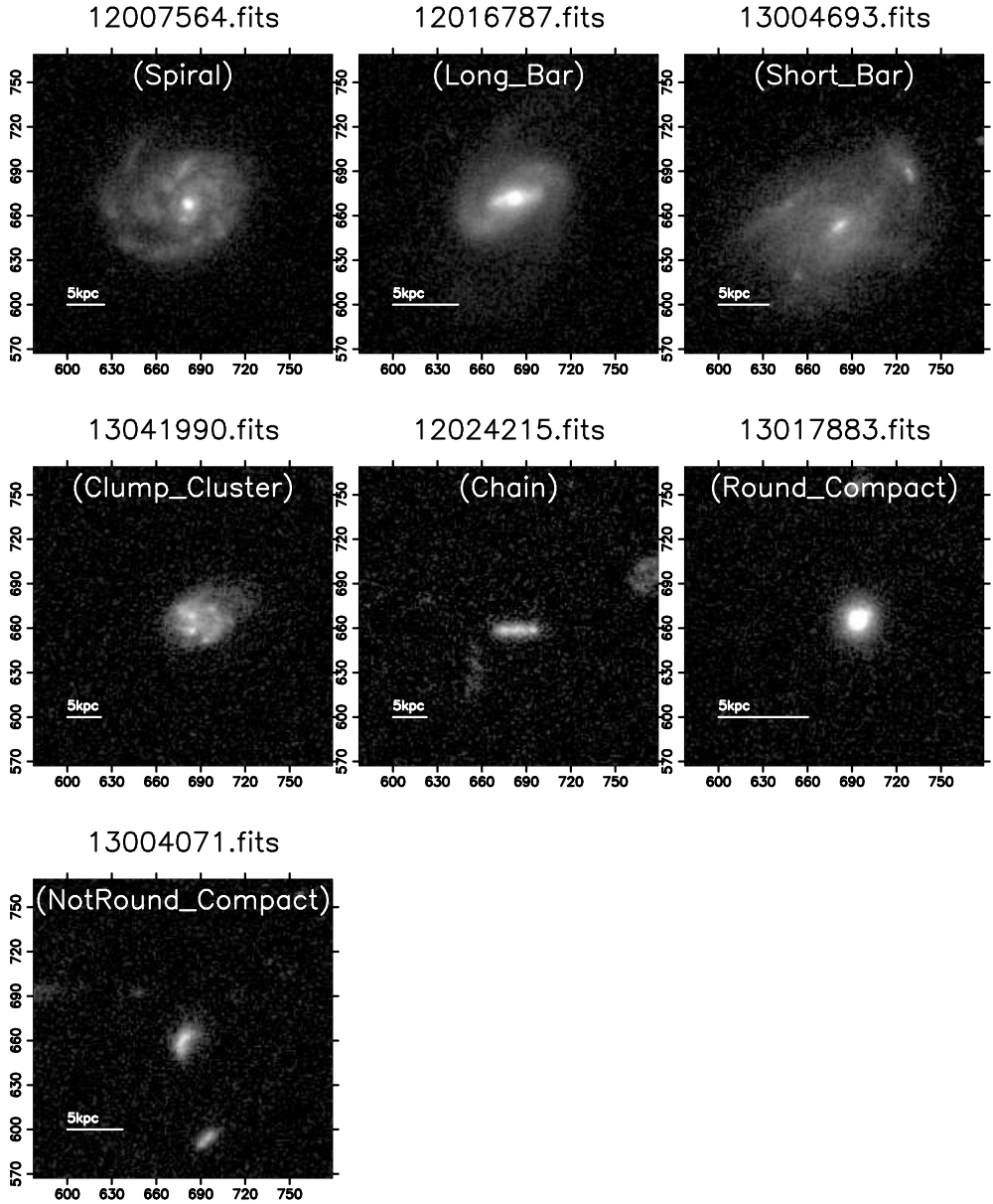}
\caption{An example for each of the classification class used to identify the galaxies from the parent DEEP2/AEGIS sample as described in \S \ref{sample}.  Each panel shows the cutout of a type of galaxy. The redshift and stellar mass are indicated at the top, the classification class is shown at the bottom with the total number in each class indicated in the parenthesis.  Also shown is a line segment for 5 kpc at that redshift and a line segment to indicate 20 pixels - the compact galaxies are usually smaller than these segments.  
} \label{classifications}
\end{figure}

We began with the 544 emission line galaxies studied by \citet{kassin07}.  For each of these galaxies, we made a cutout of the V- and I-band ACS images and visually examined each fits file.  Disk galaxies inclined more than 65$^{\circ}$, as measured by \citet{kassin07}, were discarded.    Note that an inclination cut was already made by \citet{kassin07},  removing galaxies with $i < 30^{\circ}$ and $i > 70^{\circ}$.  We also removed obviously merging galaxies and restricted the sample to z$<$0.84, following our detailed analysis of the band-shifting effect on identification of bars in \citet{sheth08}.  Beyond this redshift, the rest-frame wavelength for ACS I-band images shifts short wards of the 4000\AA\ break where bar identification becomes difficult (see Figures 7,8 and 13 in \citealt{sheth08}). We also eliminated any galaxy fainter than L$_V^*$ with an empirically determined luminosity evolution of 1 magnitude from \citet{capak03} such that M$_V^*$ = -21.7 at z=0.9 \citep{capak03}.  These criteria are based on the detailed analysis of selection effects and sample selection which are critical for high redshift studies as discussed in detail in \citet{sheth08}.  

Each of the galaxies classified  independently by four authors (DME, BGE, KS and JM) into the following classifications: barred, unbarred, clumpy (or clump-cluster), chain, and compact galaxies using postage stamps made from the optical HST data from the All-Wavelength Extended Groth Strip International Survey \citep{davis07}.  
The criteria followed were as follows: a barred galaxy was one which showed an obvious recognizable bar.  The bars were further divided into ``long'' and ``short'' bars -- ``long'' bars were those that subtended more than half of the galaxy disk.   The chain and clumpy galaxies were identified following the previous work by (e.g., \citealt{elmegreen04c, elmegreen05, elmegreen07}) - these are nascent galaxies with several bright star forming clumps believed to be in their first epoch of fragmentation and star formation. Chain galaxies are believed to be clumpy galaxies viewed edge on.  The agreement between the authors was excellent with only a 4\% disagreement between the barred versus unbarred cases.  This 4\% sample was then jointly debated and analyzed and a reconciled classification was made between all the authors.  Examples of each classification class are shown in Figure \ref{classifications}.   The final sample has 126 unbarred disk galaxies, 28 long bars and 20 short bars, 22 clumpy, 12 chain and 49 compact galaxies, for a total of 257 galaxies.  The sample is not large enough to be statistically complete or robust, as was the case for the COSMOS sample, but it is sufficient to show the basic relationship between galaxy host kinematics and development of galactic structures.   We also studied the galaxies as a function of redshift but the number of galaxies per redshift bin of $\delta$z = 0.1 was then so small ($\le$ 50) that when divided further by galaxy type, the results were not statistically meaningful.

\section{Galaxy Kinematics}

 \begin{figure}[ht!]
\plotone{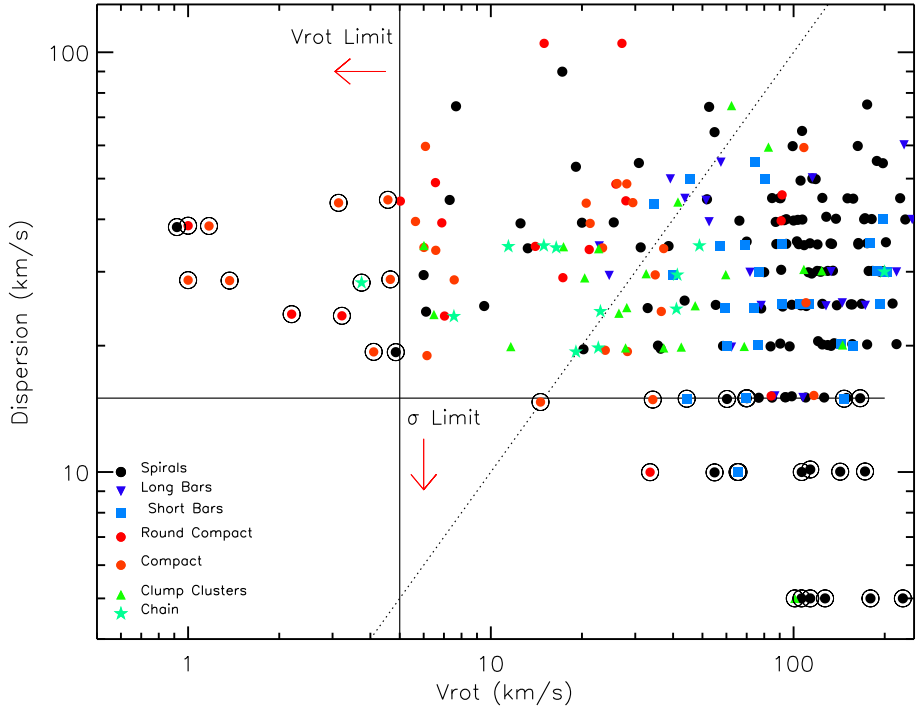}
\caption{Plot of the measured velocity dispersion versus inclination-corrected rotational velocity for different types of galaxies.  The solid black lines with red arrows indicate the limits of the measurements.  The diagonal dotted line is where the velocity dispersion and rotation velocities are equal.  Galaxies which fall below the nominal limits for these data are encircled with a black circle and are marked as such in all subsequent figures.} \label{dispvsvrot}
\end{figure}

The kinematics of the galaxies in this sample were measured from Keck DEIMOS spectra obtained by the DEEP2 Survey \citep{davis03,davis07} with the 1200 line mm$^{-1}$ grating. The kinematic measurements were first presented in \citet{kassin07}.   As the details of the observations and measurement techniques have been described elsewhere \citep{weiner06a,weiner06b,kassin07}, we simply summarize the key points here: a measure of the velocity dispersion and rotation was made from multiple bright emission-lines (typically H$\beta$, [O II] $\lambda$3727, and [O III] $\lambda$5007). These quantities were measured directly from the 2D spectral images with a routine called ROTCURVE \citep{weiner06a}. ROTCURVE constructs models of the 2D emission-line structure, convolves those models with the atmospheric seeing ($\sim0.7\arcsec$), compares to the data, and then provides a chi-squared best-fit to the values of line-intensity, velocity dispersion, and rotation. Rotational velocities, if present, were detectable down to $\sim5$ \kms, whereas dispersions were resolvable to $\sim15$ \kms (a limit set by the spectral resolution of $\sim25$ \kms). For galaxies with physical sizes smaller than the seeing, the ROTCURVE fit is a lower limit to the rotation velocity and upper limit to the dispersion.

In Figure \ref{dispvsvrot} we examine the individual measurements of the dispersion and rotational velocities as a function of galaxy type -- about one quarter of the compact galaxies have a rotation velocity measurement that is determined to be lower than the nominal 5 km/s limit for these data and should be treated with caution.  Similarly about $\sim$10\% of the unbarred disk galaxies and a handful of short bars and compact systems have a measured velocity dispersion below 15 km/s.  These systems are marked with a black circle and should be considered uncertain.  

An important caveat to note about our use of velocity dispersions from emission lines measurements is that these  may be very different from the stellar velocity dispersion because the measurements primarily reflect the mean of the local gas velocity dispersion from star forming regions.  In nearby galaxies the velocity dispersion in HII regions can easily reach several tens of km/s and the velocity dispersion from emission lines may over- or under-estimate the stellar velocity dispersion (see also \citealt{davies11}).  This important caveat however does not affect the main analysis presented here which relies on the structure of galaxies in comparison to their deviation from the Tully Fisher relation.

\section{Results} 

\begin{figure}[ht!]
\plotone{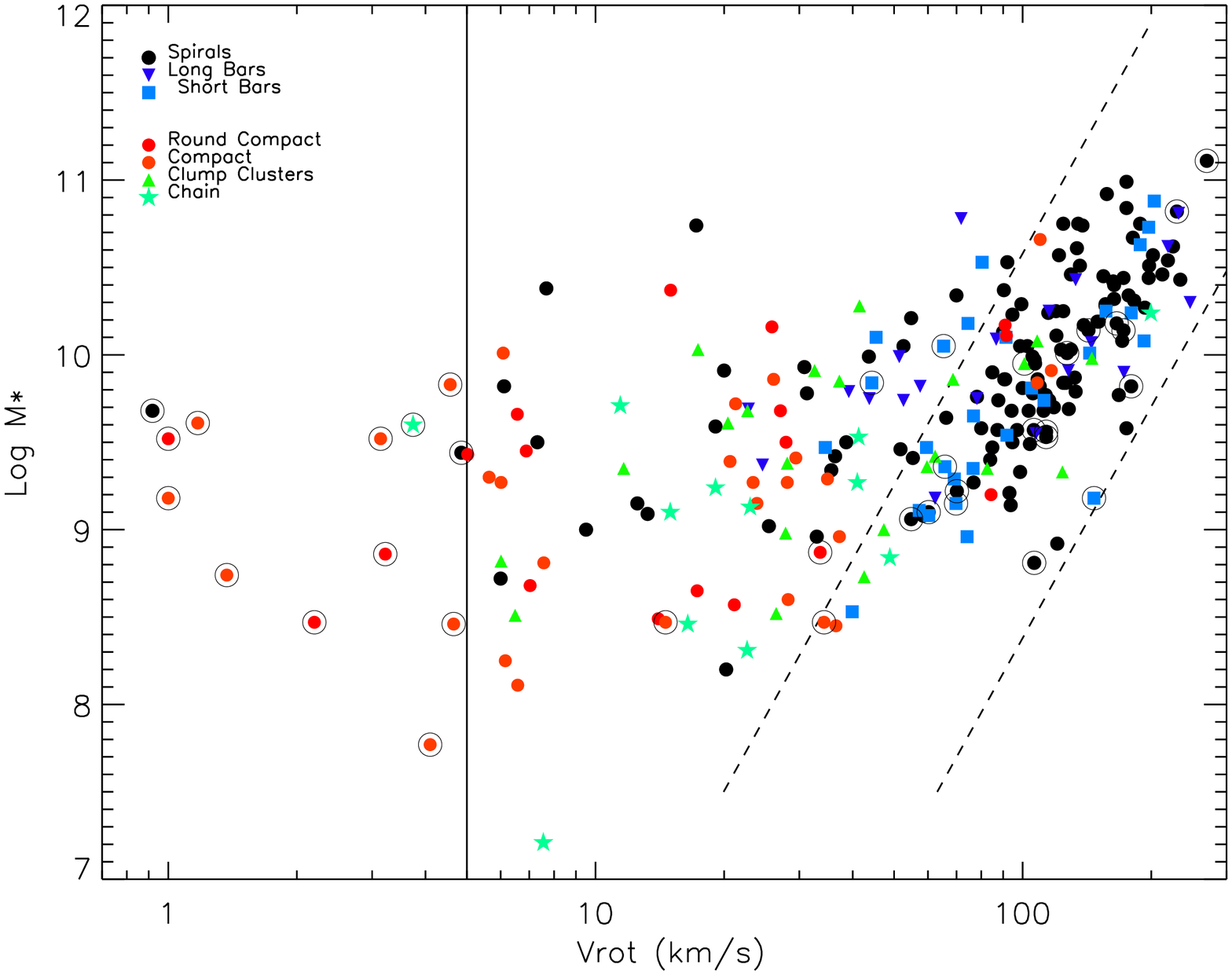}
\caption{Stellar mass and rotational velocities for the different galaxy types classifications are plotted.  The dashed lines show the width of the stellar TF-relationship between z$\sim$0 to z$\sim$1 as derived by \citet{bell01} and \citet{conselice05} respectively (and as shown in Figure 1 of \citet{kassin07}).  The vertical solid line is the same as that in Figure \ref{dispvsvrot}, showing the limits of our measurement.  The symbols for the different galaxies are as follows: dark filled circles - unbarred disks; blue triangles - long bars, light blue rectangles - short bars, filled red circles - round compact, filled orange circles - non-round compact, green triangles - clump clusters and filled stars are chain galaxies.  The black circles encircling some of the data points indicate galaxies for which the measured rotational velocity or velocity dispersion are uncertain due to the limitations of the observations.} \label{tf} 
\end{figure}

\begin{figure}[ht!]
\centering
\begin{tabular}{cc}
\epsfig{file=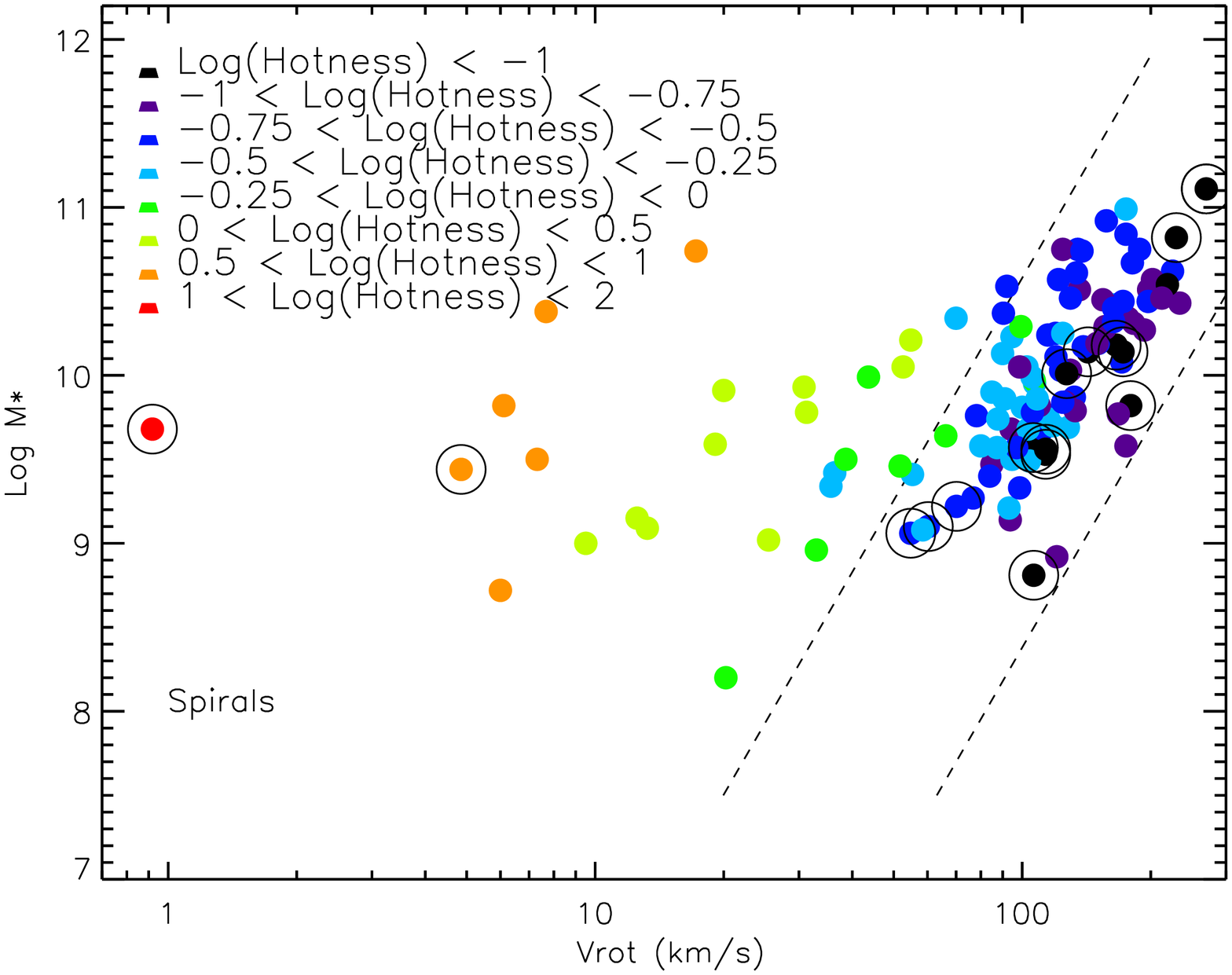,width=0.45\linewidth,clip=}   
\epsfig{file=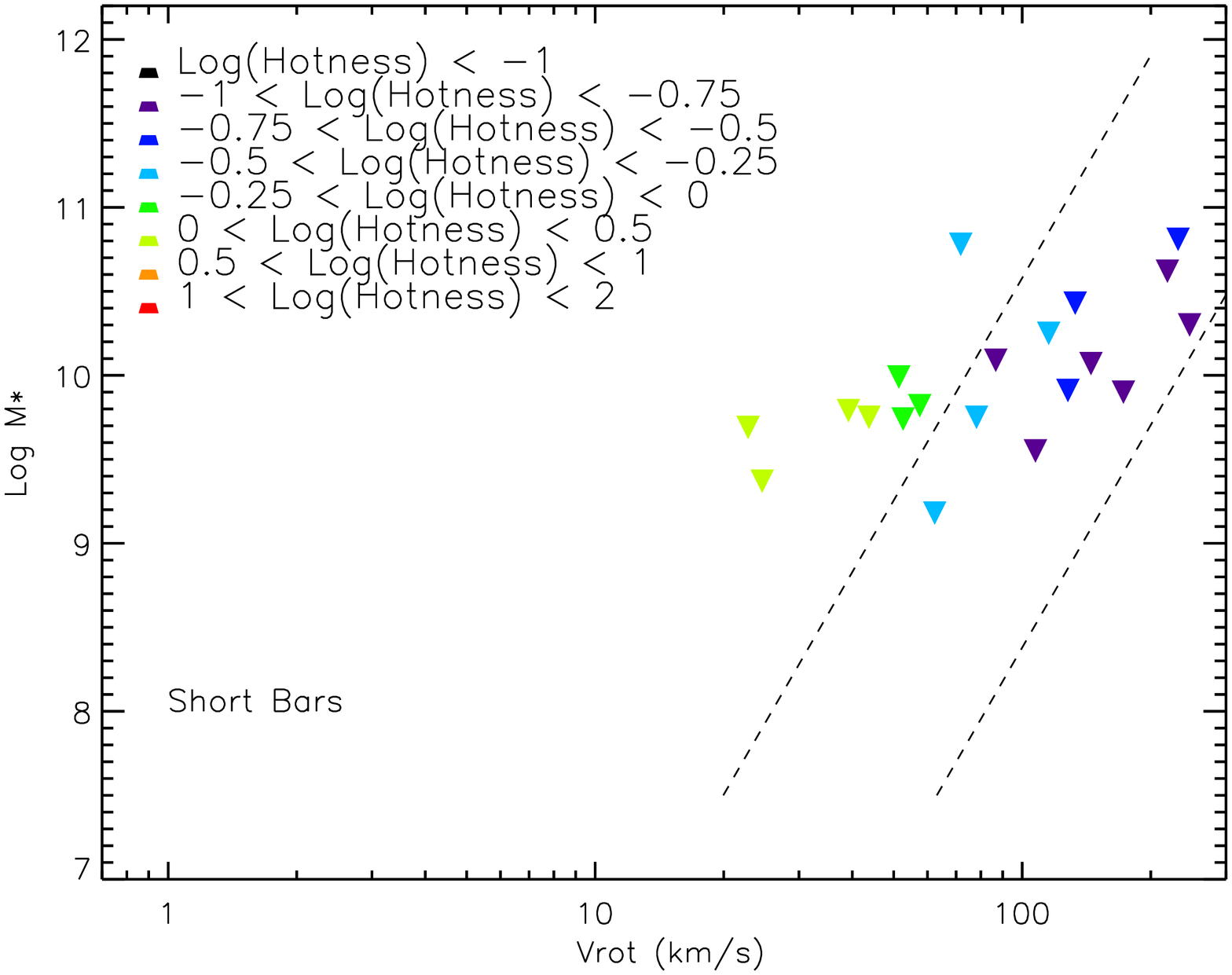,width=0.45\linewidth,clip=} \\
\epsfig{file=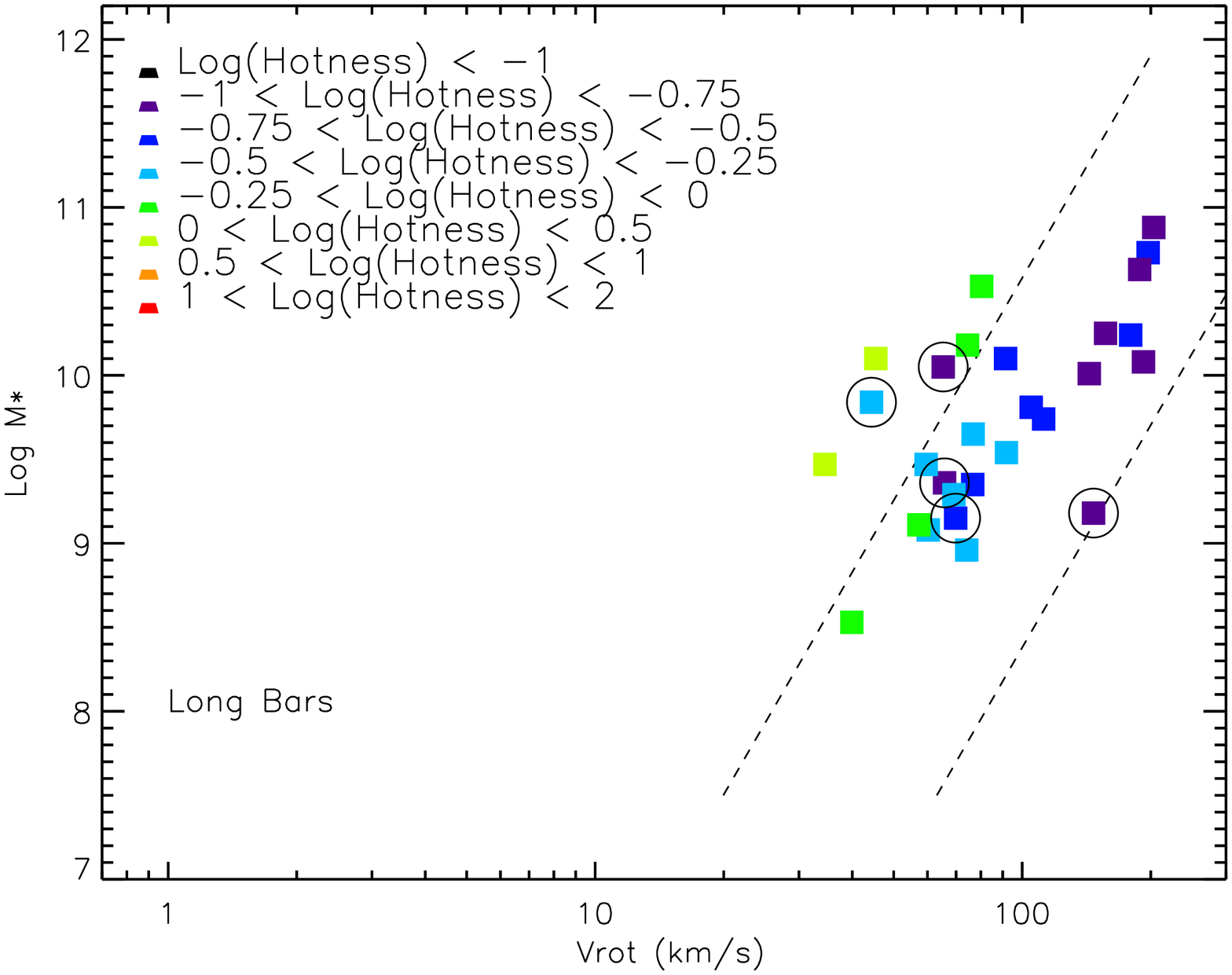,width=0.45\linewidth,clip=} 
\epsfig{file=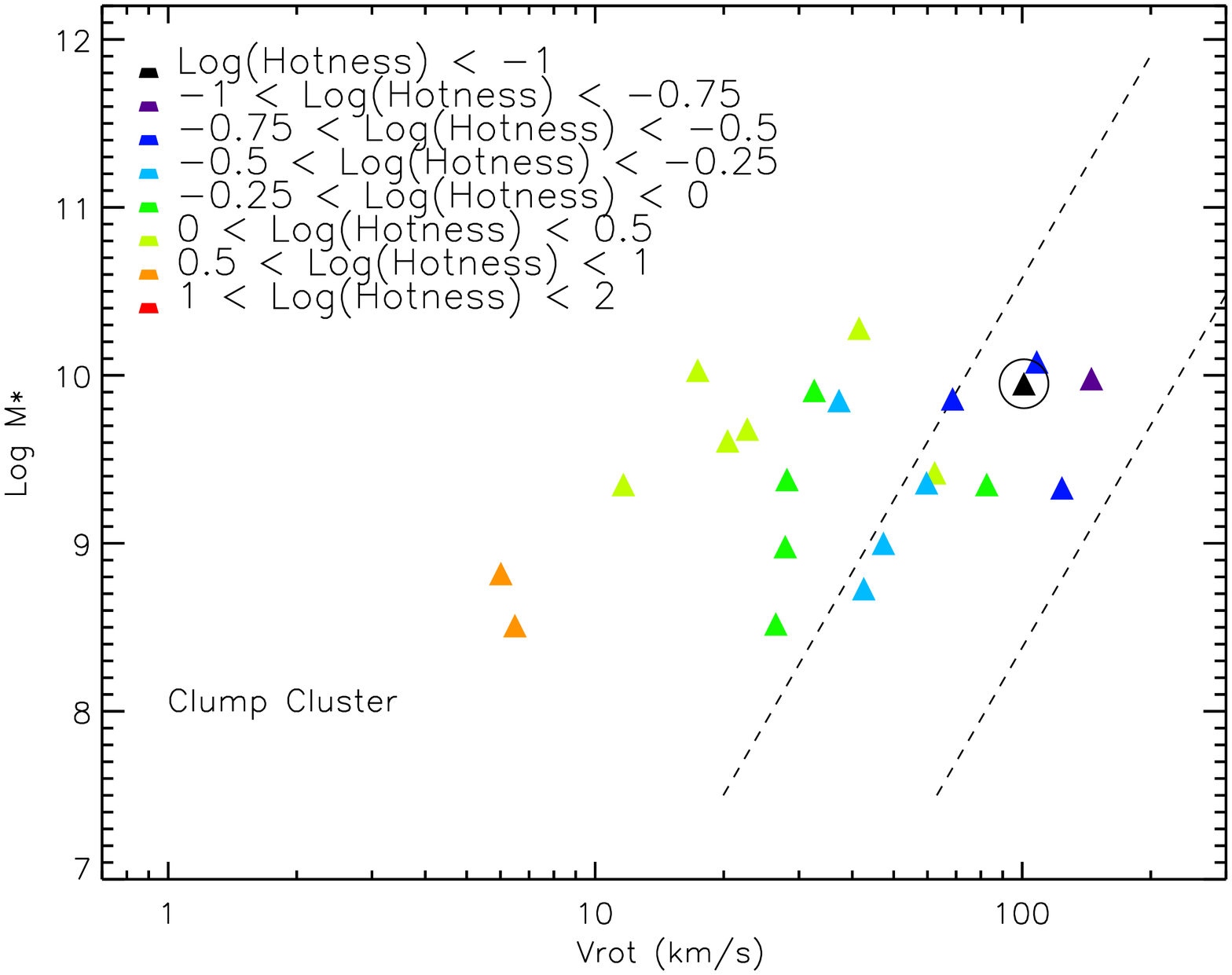,width=0.45\linewidth,clip=}  \\
\epsfig{file=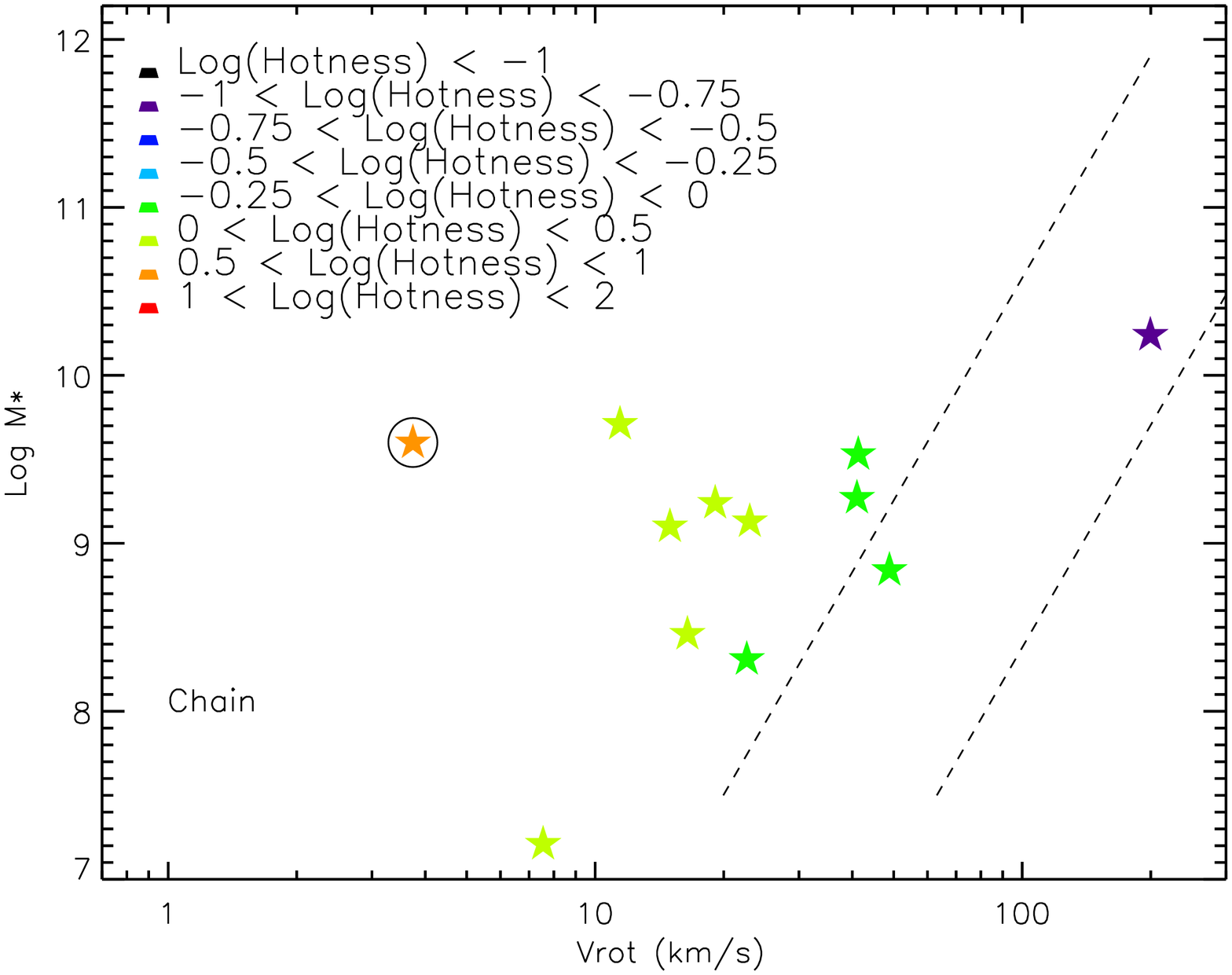,width=0.45\linewidth,clip=} 
\epsfig{file=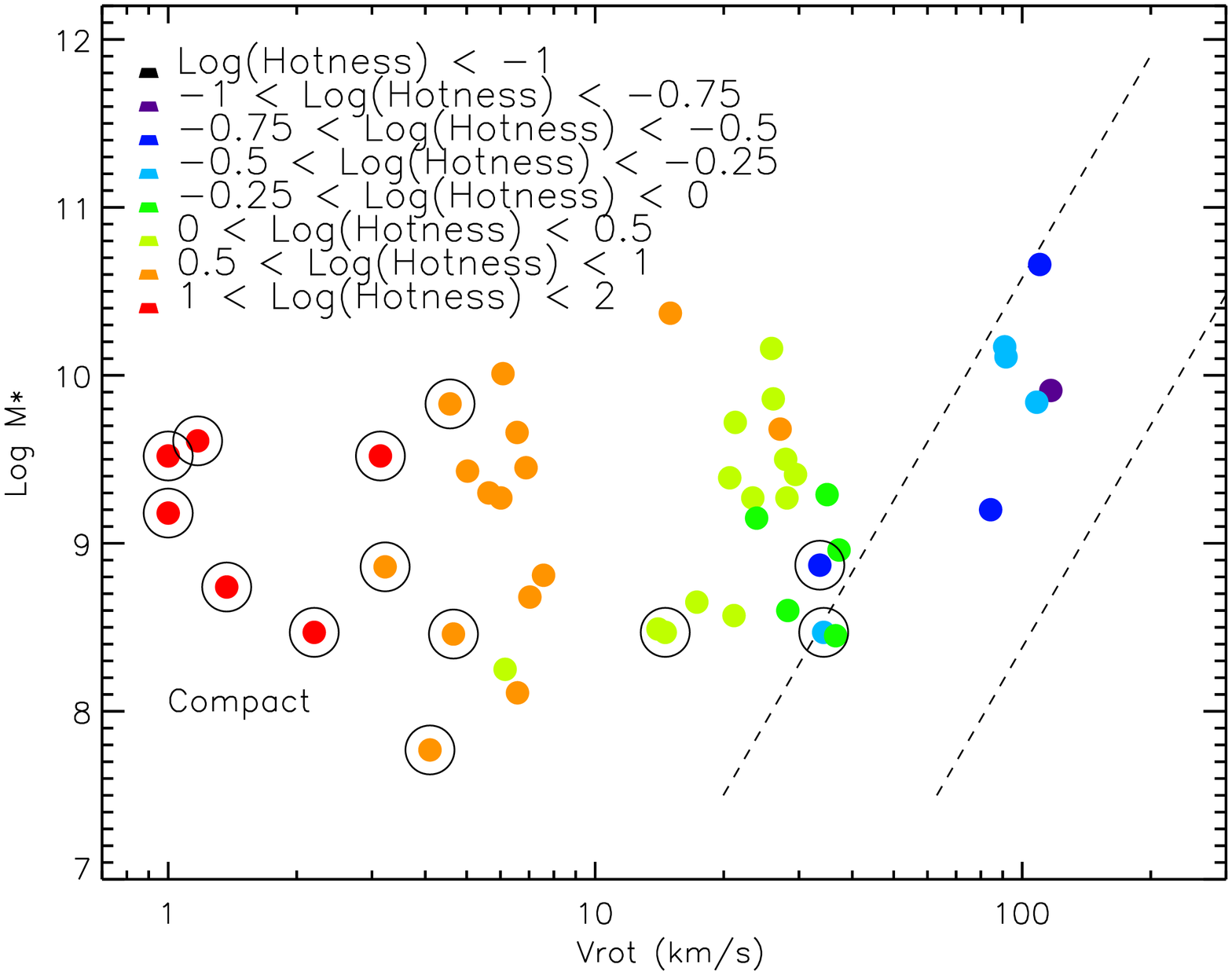,width=0.45\linewidth,clip=} 
\end{tabular}
\caption{The symbols are color coded based on the dynamical hotness of the galaxy.  The hotness bins are: log h $<$ -1  (black), -1 $<$ log h $<$ -0.75 (purple),  -0.75 $<$ log h $<$ -0.5 (dark blue), -0.5 $<$ log h $<$ -0.25 (light blue), -0.25 $<$ log h $<$ 0 (dark green), 0 $<$ log h $<$ 0.5 (light green), 0.5 $<$ log h $<$ 1 (orange), and 1 $<$ log h $<$ 2 (red).  The location of the galaxies on the stellar TF is {\em strongly} correlated with the hotness of the disk. There is a gradual progression from dispersion-dominated systems to rotationally dominated systems from the left to right in every type of galaxy, including within the barred galaxies.} \label{tfwhot}
\end{figure}

We combine the data from the \citet{kassin07} study with our classifications described above.  The results are plotted in Figure \ref{tf} and Figure \ref{tfwhot}.  These two figures show the main result of this study -- barred spirals (blue squares and purple triangles) in this high redshift sample are primarily found in massive, rotationally-supported galaxies that are on the TF, as is the case for galaxies in the local Universe \citep{courteau03}.  There is a large ``tail'' of galaxies to the lower rotational velocities primarily made of compact (red or orange circles) and clumpy/chain (green triangles/stars) systems.  There are a few ($\sim$20/126 or 15\%) unbarred disk galaxies in the region off the TF but there are virtually no barred spirals far from the envelope of the TF.  

In Figure \ref{tfwhot} we plot the TF shown in Figure \ref{tf}  but now color coded with ``dynamic hotness'', i.e. using the ratio of the velocity dispersion to the rotational velocity (h = $\sigma$ / V$_{rot}$) and separated by galaxy type in each of the six panels.  Galaxies with log h $>$ 0 are dispersion-dominated whereas those with log h $<$0 are rotation-dominated.  As noted by \citep{kassin07}, there is  a clear trend with galaxies migrating to the TF with decreasing hotness.   As would be expected, the rotation-dominated systems (blue--black colors) are on the TF and are primarily made of unbarred and barred spirals.  About half of the clumpy galaxies and a few compact systems are also rotation-dominated whereas the dispersion-dominated systems are predominantly composed of compact galaxies.  One difference to note is the dynamic state of the barred population -- $\sim$20\%  of the long bars are coded light or dark green, indicating -0.25 $<$log h $<$ 0.5, compared to 40\% of the short bars.  Although the sample sizes are relatively small, the difference is suggestive that short bars are preferentially in somewhat hotter disks than long bars.  

\section{Discussion}

The main result of this paper is that bars are not present in dispersion-dominated disk galaxies.  The data suggest an evolutionary sequence in the assembly of disks and formation of the familiar galactic structures such as bars that we see today.   The clump-cluster and chain galaxies are believed to be an early phase of present-day spiral galaxies undergoing a burst of star formation in large, gravitationally unstable clumps in a cold, gaseous disk (e.g., \citealt{elmegreen05,elmegreen09}).  As these disks evolve and accrete more cold material from the large scale structure filaments, they should evolve towards more rotationally-supported disks, the type that we see on the Tully-Fisher relationship (see also discussion in \citealt{kassin07}).  While it is not clear how these systems migrate from the left hand side of the diagram to the right, the expected evolution is likely to be towards higher rotational velocities and stellar masses (up and to the right) on Figures \ref{tf}.   The data also suggest that bars may be growing from short to long as the disks evolve to colder and more massive systems, indicating that bar-driven heating of the disk is less significant than the competing cooling processes.

\begin{figure}[ht!]
\centering
\begin{tabular}{ll}
\plotone{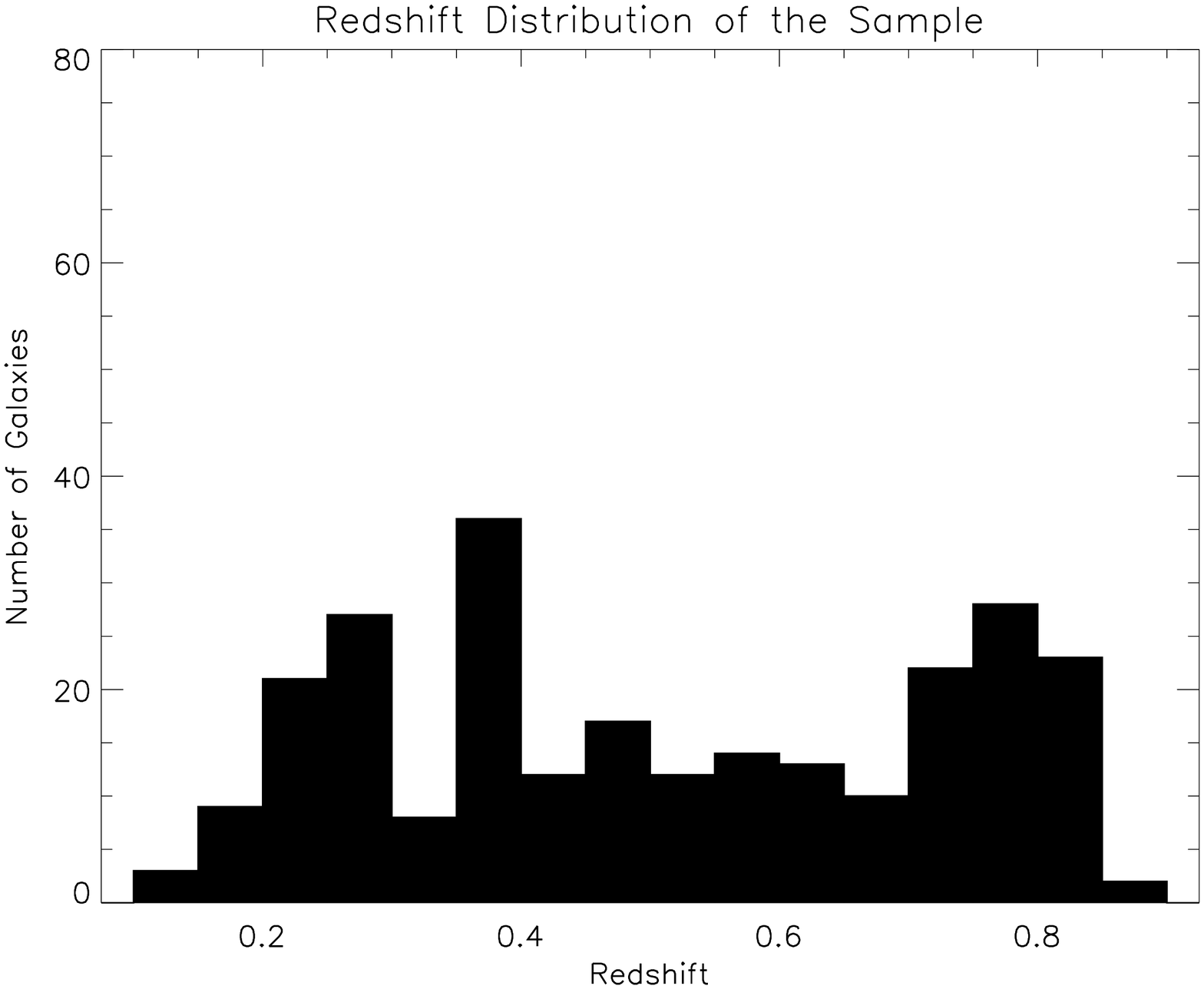}   
\end{tabular}
\caption{Redshift distribution for the 257 galaxies in this sample is shown here.  The median redshift of the sample is 0.46.  There are too few galaxies in each redshift bin to discuss evolutionary trends with both mass and redshift which will have to wait for even larger surveys.} \label{histz}
\end{figure}

Previous studies have shown a strong correlation between the bar fraction, stellar mass of the galaxy and redshift such that massive galaxies ($>$ 10$^{11}$ \Msun) had a high ($>$50\%) bar fraction at z$\sim$0.85, whereas lower mass galaxies (10$^{10}$\Msun) had a bar fraction $<$ 20\% \citep{sheth08}. The evolution of the bar fraction was differential over the last 7 Gyr with the fastest growth of the bar fraction occurring in the low mass, blue, late type spirals of masses between 10$^{10}$ and 10$^{11}$ \Msun \citep{sheth08,cameron10}.  The present DEEP2/AEGIS sample is too small to measure the redshift and mass dependent evolution of these galaxies.  The redshift distribution is shown in Figure \ref{histz}. While we do see a segregation along the stellar mass axis between compact galaxies and disk (barred and unbarred) galaxies there are too few galaxies to infer any trends with mass and redshift - further analysis with larger data sets will be very useful to interpret the evolutionary trends with mass and redshift.  Finally the compact systems, which are primarily dispersion-dominated, are seen over the entire redshift range of this survey and are therefore not necessarily only exotic high redshift systems, as was found in \citet{kassin07}.  The fate of these objects is another interesting area of study, especially at lower redshifts, where high spatial resolution (and higher signal to noise) are available.

Although our data shed some light on the conditions that delay bar formation, the large number of non-barred galaxies that are massive, cold and rotationally supported remains a mystery.  \citet{courteau03} have already shown that there is no obvious difference in the placement of barred and unbarred spirals on the TF in the local Universe.  Locally as many as 30-35\% of the disk galaxies are unbarred.  And so we conclude that while dynamic coldness and sufficient stellar mass are necessary conditions for the formation of a bar, they are not sufficient. Mergers and interactions are other processes  which could play a role in bar formation but their impact is difficult to quantify because they can create long-lived or  transient bars or they can destroy existing bars \citep{gerin90,barnes91,mihos94,mihos96,romanodiaz08}).   Finally, another important process for bar formation is the interaction history between the baryonic matter and the dark matter halo, especially in the inner parts of galaxies because the dark matter halo can act as sink of angular momentum and energy for the baryonic matter settling into the central bar (e.g., \citet{athanassoula02}).

\acknowledgments
\section{Acknowledgments}
\acknowledgments
We thank the referee for the careful reading and suggestions that improved the paper.  We are also grateful to the DEEP2 team for graciously sharing their data for this analysis.  We thank Richard Ellis, Rick Fisher, Sarah Miller and Preethi Nair for insightful discussions that helped improve this paper.  KS acknowledges support from the National Radio Astronomy Observatory is a facility of the National Science Foundation operated under cooperative agreement by Associated Universities, Inc.  EA acknowledges financial support to the DAGAL network from the People Programme (Marie Curie Actuins( if the European Union's Seventh Framework Programme FP7/2007-2013/ under REA grant agreement number PITN-GA-2011-289313.  
{\it Facilities:} \facility{{\em Hubble} Space Telescope}

\bibliography{mymasterbib}  

\end{document}